\documentclass[graybox]{svmult}

\usepackage{mathptmx}       
\usepackage{helvet}         
\usepackage{courier}        
\usepackage{type1cm}        
\usepackage{makeidx}         
\usepackage{graphicx}        
\usepackage{multicol}        
\usepackage[bottom]{footmisc}

\makeindex             

\begin{document}

\title*{Variability survey in the young open cluster NGC\,457}
\author{Dawid Mo\'zdzierski, Andrzej Pigulski, Grzegorz Kopacki, Marek St\c{e}\'slicki and Zbigniew Ko{\l}aczkowski}
\authorrunning{Mo\'zdzierski et al.}
\institute{D.\,Mo\'zdzierski \at Instytut Astronomiczny, Uniwersytet Wroc{\l}awski, Kopernika 11, 51-622, Wroc{\l}aw, Poland, \email{mozdzierski@astro.uni.wroc.pl}}
%
%
\maketitle

\abstract*{We present preliminary results of the photometric variability search in the field of view of the young open cluster 
NGC\,457. We find over 60 variable stars
in the field, including 25 pulsating or candidate pulsating stars.}

\abstract{We present preliminary results of the photometric variability search in the field of view of the young open cluster 
NGC\,457. We find over 60 variable stars
in the field, including 25 pulsating or candidate pulsating stars. }

\section{The cluster}
\label{sec:1}
NGC 457 is a young open cluster in Cassiopeia, nearby $\phi$ Cas. Its age is estimated for 10--20 Myr \cite{PheJan1994,Loktin2001}, 
distance, for about 2.5--3.0 kpc. It is located in the Perseus arm of the Galaxy.
Six variable or suspected variable stars were known in the observed cluster field prior to our study (see, e.g., \cite{Maciejewski2008}). We
present preliminary results of the photometric variability survey in this cluster aimed at discovery of B-type
pulsating stars and bright eclipsing binaries.

\section{Observations and results}
\label{sec:2}
The photometric observations of NGC 457 were obtained during three runs: the run consisting of 4 nights in
1993 carried out in Ostrowik station, University of Warsaw, by one of us (GK), the second run of 31 nights in the years 1999--2002 
(Bia{\l}k\'ow station, University of Wroc{\l}aw) and
the third one consisting of 24 nights made again in Bia{\l}k\'ow between December 2010 and March 2011. Here, we
present results based on the 2010--2011 observations only. Of the three runs, the last one is of the best quality.
During this run we used 60-cm reflecting telescope equipped with the Andor Tech.~DW 432-BV
CCD camera covering 13$^{\prime}\times$\,12$^{\prime}$ field of view. All frames were calibrated in a standard way and reduced
with the Daophot II package \cite{Stetson1987}.

\begin{figure}[!ht]
\sidecaption
\includegraphics[scale=.65]{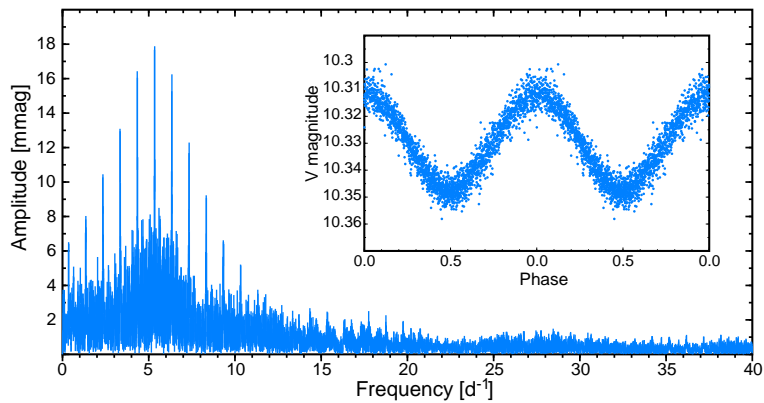}
%
%
\caption{Fourier amplitude spectrum for the $V$-filter data of $\beta$~Cephei star NGC 457-8. 
The inset shows the light curve phased with the main period.}
\label{fig:1}  
\end{figure}

We have found 64 variable stars in the observed field, including the six already known. The most interesting is the discovery of
pulsating stars, likely members of the cluster. The sample of pulsating stars
includes: a single $\beta$~Cephei star NGC 457-8 (see Fig.~\ref{fig:1}), 12 (candidate) SPB stars and 12 $\delta$~Scuti stars. 
In addition, nine eclipsing and ellipsoidal variables were found. 
The sample includes also stars showing irregular or (quasi)periodic variations of
unknown origin both in the cluster main sequence and among the reddest stars in the field. Some of
B-type stars showing this type of variability are known as Be stars. 

%

\begin{acknowledgement}
We thank Dominik Drobek, Piotr \'Sr\'odka and Ewa Zahajkiewicz for making some observations of NGC\,457. The work was
supported by the MNiSzW grant No. N N203 302635.
\end{acknowledgement}

\end{document}